\pdfoutput=1

\documentclass[%
 reprint,
 amsmath,amssymb,
 aps, float,
prb,
floatfix
]{revtex4-2}

\usepackage{graphicx, ulem, color, soul, tabularx, textcomp}
\usepackage{dcolumn}
\usepackage{bm, xcolor, siunitx, float}
\begin{document}

\preprint{APS/123-QED}

\title{Tripling of the Superconducting Critical Current Density in BaFe$_2$(As$_{1-x}$P$_x$)$_2$ Retained After Pressure Release}

\author{Jiangteng Liu}
\affiliation{Department of Electrical \& Computer Engineering,
University of Washington, Seattle, Washington 98195}

\author{Alex Lopez}
\affiliation{Department of Mechanical Engineering,
University of California, Los Angeles, Los Angeles, California 90095, USA}

\author{Daniel Slone}
\affiliation{Department of Materials Science \& Engineering,
University of Washington, Seattle, Washington 98195, USA}

\author{Guodong Ren}
\affiliation{Department of Materials Science \& Engineering,
University of Washington, Seattle, Washington 98195, USA}

\author{Zhaoyu Liu}
\affiliation{Department of Physics,
University of Washington, Seattle, Washington 98195, USA}

\author{Juan-Carlos Idrobo}
\affiliation{Department of Materials Science \& Engineering,
University of Washington, Seattle, Washington 98195, USA}
\affiliation{Physical Sciences Division,
Pacific Northwest National Laboratory, Richland, Washington 99352, USA}

\author{Jiun-Haw Chu}
\affiliation{Department of Physics,
University of Washington, Seattle, Washington 98195}

\author{Serena Eley}
\email{serename@uw.edu}
\affiliation{Department of Electrical \& Computer Engineering,
University of Washington, Seattle, Washington 98195, USA}

\date{\today}

\begin{abstract}

Superconducting performance is tunable not only via chemical modification or defect engineering, but also through external parameters such as pressure, though this method remains less readily accessible. In this work, we study how compression influences vortex dynamics and critical currents in an iron-based superconductor. Specifically, we perform magnetization measurements using an off-the-shelf pressure cell to investigate the effects of hydrostatic pressures up to 1.08 GPa on the magnetic properties of BaFe$_2$(As$_{0.62}$P$_{0.38}$)$_2$ crystals across a range of temperatures $T$ and magnetic fields $H$. Although these pressures minimally affect the superconducting critical temperature, they produce a clear increase in the critical current density $J_c(T,H)$, a pronounced reduction in the rate of thermally activated vortex motion $S(T,H)$, and can change the dominant vortex pinning mechanism. Furthermore, the effects of pressure are irreversible: after pressurization and subsequent release at room temperature, high-density microcracks are observed and the crystals retain their enhanced critical current densities. The second magnetization peak vanishes above 18 K after the pressure cycle, which we attribute to a transition from predominantly $\delta \kappa$ pinning to a mixed mechanism of $\delta T_c$ and surface pinning. Lastly, a threefold increase in $J_c$, a more than 40\% reduction in $S$ at 8~K and 0.5~T, and an expanded elastic-creep region were achieved after $1-2$ pressure cycles. These findings demonstrate the potential utility of pressure cycling for improving $J_c$, which may offer a simpler alternative compared to approaches such as chemical doping or the introduction of artificial pinning centers.

\end{abstract}

\maketitle

\section{\label{sec:introduction} Introduction}

Pressure is an effective tool for tuning superconductivity, with the potential to enhance transition temperatures, alter orbital overlap, adjust electron‑phonon coupling strength, suppress competing magnetic orders, and induce structural transitions that give rise to new phases \cite{PhysRevB.89.220502, 9qx8-7q4x, Lorenz2005, Sang2021, Yu2021, PhysRevB.111.L081110, Gati2020, PhysRevX.6.041045, Hung2021}. In many systems, applied pressure mimics the effects of chemical doping, as both strain the lattice and alter electronic structure, providing a clean, potentially reversible route to probe structure–property relationships without introducing chemical disorder. 

Even modest compressive strains of a few gigapascals may strongly enhance superconductivity. For example, in FeSe applying only
1.48 GPa drives the $T_c$ from 9 K to 27 K \cite{Mizuguchi2008} possibly due to enhanced antiferromagnetic spin fluctuations \cite{PhysRevLett.102.177005}. Importantly, superconducting performance in applications is governed not only by transition temperature but also by the motion of vortices --- penetrating magnetic flux lines whose dynamics directly determine the critical current density $J_c$. Defects and doping can pin vortices, suppressing their motion and thereby enhancing $J_c$, making vortex physics central to both fundamental understanding and technological optimization. 

In fact, pressure has been shown to improve vortex pinning and the critical current density \cite{Shabbir2015, Shabbir2016}, for example, in Fe$_{1-x}$Co$_x$Se$_{0.5}$Te$_{0.5}$ \cite{Sang2018,sang2019}. Despite these promising effects, maintaining high pressure during normal operation is impractical for most real-world applications, motivating strategies that can preserve pressure-induced improvements after pressure release. For example, recent interest has arisen in the pressure‑quench protocol (PQP), in which superconductors retain pressure‑induced structural modifications after pressure release, potentially stabilizing enhanced states at ambient conditions \cite{Deng2021,Deng2025}. 

In this work, we employ an off‑the‑shelf pressure cell to study iron‑based superconductors, focusing on two questions: how applied pressure influences $J_c$ and thermally activated vortex motion (creep parameter $S$), and whether these crystals can preserve pressure‑induced enhancements after release. This approach probes the role of lattice strain in vortex dynamics and superconducting performance, while assessing the practicality of using a compact, off-the-shelf pressure cell and room‑temperature release as a pathway toward engineering improved superconductors.

\section{Results and Discussion}

We studied the effects of hydrostatic pressure on two overdoped, superconducting BaFe$_2$(As$_{0.62}$P$_{0.38}$)$_2$ (P-doped Ba122) single crystals (Sample \#1 and \#2) grown using a self-flux technique. In particular, we conducted microscopy, determined the pressure dependence of the critical current density $J_c(T,H)$ and rate of thermally activated vortex motion $S(T,H)$ from magnetization studies conducted over a range of temperatures $T$ and magnetic fields $H$. Because magnetometry also detects the signal from the sample mount --- which is especially substantial for a pressure cell --- meticulous background subtraction protocols, described below, are requisite for accurate results. Details of the crystal growth, magnetometry measurements, and electron imaging setup are provided in the Methods section.

\subsection*{Pressure cell characterization}

\begin{figure*}[!ht]
\centering
\includegraphics[width=1\linewidth]{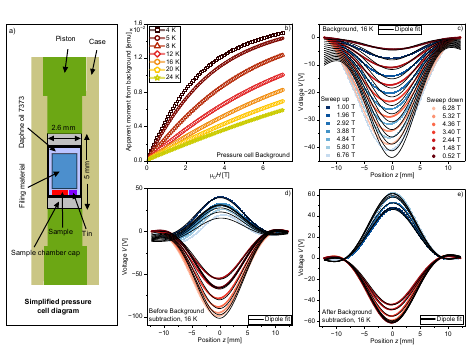}
\caption{\label{fig:fig1}(a) Schematic of the Quantum Design MPMS 3 pressure cell. (b) Field-dependent background moment of the pressure cell measured at different temperatures. The “apparent moment” refers to the moment parameter returned by the dipole fit applied to a non-dipolar background response. (c–e) Selected voltage vs position raw scans at 16 K for sweep up (blue square) and down (red circle) ($0-7-0$ T) the magnetic field. The black curve indicates the dipole fit. (c) Pure background signal. (d) Combined background and sample signal before background subtraction. (e) Sample signal after background subtraction. It is important to note that we added a constant voltage offset $V_0$ to each scan, so that the voltage begins around zero, to reduce overlap between curves in the plots. The offset $V_0$ does not affect the extracted magnetic moment, since the moment depends only on the relative dipole amplitude $A$, as described in Eq.~(\ref{eq:dipole}).
}
\end{figure*}

All measurements of the magnetic moment $m$ were collected using a Quantum Design MPMS3 magnetometer, with the sample mounted either on a standard brass sample holder or inside an off-the-shelf 1.3 GPa model Quantum Design BeCu pressure cell. For pressure-dependent measurements, the sample was loaded into a Teflon capsule with an outer diameter of 2.6 mm and an inner diameter of 2.2 mm, as illustrated in Fig.~\ref{fig:fig1}(a). The pressure was calibrated using a tin (Sn) manometer, as the critical temperature $T_c$ of pure Sn exhibits a linear pressure dependence of $dP/dT_c = -0.489~\mathrm{GPa/K}$ in the range $0-1.3$~GPa. The tin piece was placed directly adjacent to the sample to ensure that the measured pressure accurately reflects the pressure experienced by the crystal. The pressure-dependent $T_c$ of the tin manometer is shown in Supplementary Fig. S1. 

The pressure cell introduces a large background signal (e.g., 0.015 emu at 4 K and 7 T) that varies strongly with both temperature and magnetic field, as shown in Fig.~\ref{fig:fig1}(b). In particular, the field dependence becomes markedly nonlinear at low temperatures. These complex features make accurate background subtraction essential to determine the true sample moment. Note that the background shows no time dependence, as demonstrated in Supplementary Fig.~S6.

The Quantum Design MPMS-3 software \textit{MultiVu} determines the magnetic moment by measuring the voltage $V(z)$ as the sample is scanned through a second-order gradiometer pickup coil set, where $z$ denotes the sample position relative to the center of the coil assembly. The gradiometer consists of four pickup coils: two closely spaced central coils and two outer coils located symmetrically at each end. These coils form a closed superconducting loop that is coupled to a superconducting quantum interference device (SQUID) sensor. The specific shape of $V(z)$ is determined by the geometry of the second-order gradiometer. Here, the top and bottom coils are wound in the opposite direction relative to the two central coils. When the sample moves through the coil set, the flux coupled into coils with opposite winding directions changes sign, resulting in a voltage signal that first decreases, then increases, and then decreases again for a positive magnetic moment. If the magnetic moment reverses sign, the magnetic field distribution is inverted, leading to an overall sign reversal of the entire $V(z)$ signal. Next, the signal is then fit to the theoretical dipole response function \cite{QD1014_213,QD1500_022,QD1500_023,SQUIDLAB, Randy_SQUID_2021}:

\begin{align}
V(z) = A\left\{
\frac{2}{(R^{2} + z^{2})^{3/2}}
- \frac{1}{[R^{2} + (z+L)^{2}]^{3/2}} \right. \nonumber \\
\left. - \frac{1}{[R^{2} + (z-L)^{2}]^{3/2}}
\right\} + V_0
\label{eq:dipole}
\end{align}

\noindent Here, $R$ is the gradiometer radius, $L$ is the gradiometer half length, and $V_0$ is the voltage offset. The resulting fitted dipole amplitude $A$ is then converted to a magnetic moment via a fixed calibration constant.  Therefore, background subtraction on the extracted moment can only be performed if the background itself follows a dipolar response. However, as shown in Fig.~\ref{fig:fig1}(c), the magnetic background of the pressure cell does not exhibit a dipole-like shape, and thus the subtraction must instead be carried out at the raw voltage–position stage. Figure~\ref{fig:fig1}(d) presents the voltage–position scans of the sample prior to background subtraction. This data was collected at 16 K with the magnetic field swept up and down (0~T to 7~T to 0~T). The dipole fits (black curves) poorly match the data. By comparison, Fig.~\ref{fig:fig1}(e) shows the corrected scans after background subtraction, where a reliable dipole fit can be achieved, yielding an accurate magnetic moment for the sample. It is important to note that when the background signal becomes larger than the sample signal, background subtraction may become unreliable. Further details regarding the fitting robustness are provided in Supplementary section 5, titled “Magnetic background of the 1.3 GPa Model high pressure cell”. In addition, when the sample volume is less than 60--70 \% of the total volume, a filler material is required to achieve the designed pressure. In our measurements, we used a Teflon spacer as the filler. This configuration may introduce a non-uniform pressure environment during the measurement.

\subsection*{Effect of pressurization on vortex pinning}

\begin{figure}[!ht]
\centering
\includegraphics[width=1\linewidth]{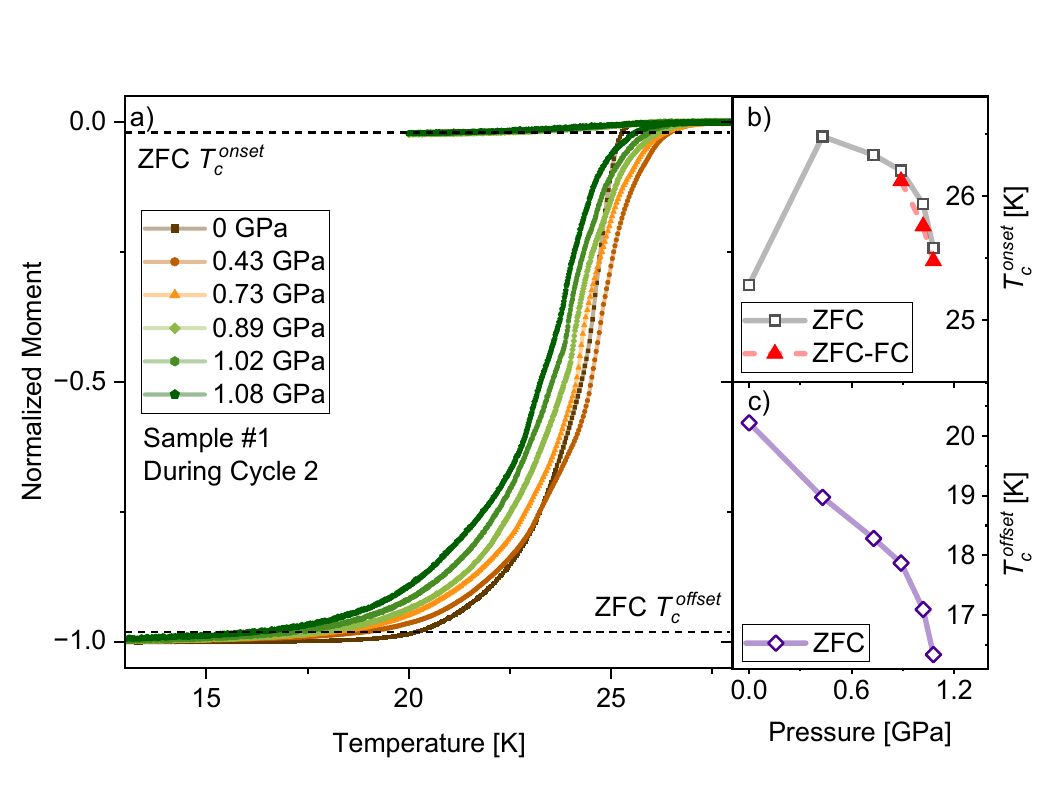}
\caption{\label{fig:fig2} (a) Normalized magnetic moment as a function of temperature for Sample \#1 during the second pressurization cycle. Here, an applied field of $H = 10$ Oe was used at 0 GPa, while $H = 3$ Oe was used for all subsequent pressures. (b) Pressure dependence of $T^{onset}_c$, where $T^{onset}_c$ values were extracted using two approaches: ZFC-only, and the difference between the ZFC and FC curves. (c) Pressure dependence of $T^{offset}_c$. }
\end{figure}

To investigate the effect of hydrostatic pressure on the superconducting state of overdoped BaFe$_2$(As$_{1-x}$P$_x$)$_2$, we measured the magnetic moment against temperature under pressures ranging from 0 to 1.08~GPa in both zero field cooling (ZFC) and field cooling (FC) conditions, as shown in Fig.~\ref{fig:fig2}(a). With increasing pressure, the superconducting transition shifts progressively to lower temperatures. Because this shift is relatively small, we extracted $T_c$ using multiple criteria to ensure that the trend is reliable.

The pressure dependence of the onset transition temperature $T_c^{\mathrm{onset}}$ is summarized in Fig.~\ref{fig:fig2}(b). We used two methods to determine $T_c^{\mathrm{onset}}$:  
(1) a ZFC criterion, defined as the temperature where the normalized moment decreases below $-0.02$, and  
(2) a ZFC-FC criterion, defined as the temperature at which the difference between the ZFC and FC curves exceeds 0.02 of the normalized moment. Apart from this, we also extracted the offset transition temperature $T_c^{\mathrm{offset}}$, defined as the temperature where the normalized moment falls below $-0.98$, as shown in Fig.~\ref{fig:fig2}(c). As expected for overdoped compositions, the extracted $T_c$ decreases with increasing pressure~\cite{Klintberg2010, Goh2010}. The only deviation occurs for $T^{onset}_c$ at the 0~GPa point. We speculate that this behavior arises from non-homogeneous pressure (from the filler material) at 0.43 GPa and above. To quantify this, the transition width ($T^{onset}_c - T^{offset}_c$) against pressure is plotted in Supplementary Fig.~S4(c). This can also explain the relatively broad transition evident in Fig~\ref{fig:fig2}(a).

We next examined the temperature-dependent critical current density $J_c(T)$ under different applied pressures. Here, $J_c$ was calculated using the Bean critical state model for rectangular samples \cite{Gyorgy1989, Talantsev2024BeanModel},
\begin{equation}\label{eq:Bean}
    J_c(T) = \frac{20\,\Delta m(T)}{w^2 l\, \delta \left(1 - \frac{w}{3l}\right)},
\end{equation}
where $\Delta m$ is the difference in moment between the upper and lower branches of the hysteresis loop, and $w$, $l$, and $\delta$ denote the width, length, and thickness of the crystal, respectively. Here, $J_c$, $m$, the geometric dimensions, and the prefactor $20$ carry units of A\,cm$^{-2}$, emu, cm, and A\,cm$^{-2}$\,emu$^{-1}$, respectively.

\begin{figure}[!ht]
\centering
\includegraphics[width=1\linewidth]{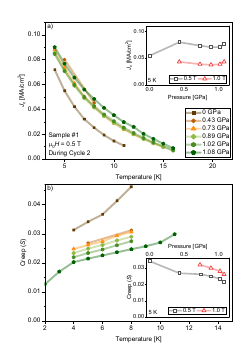}
\caption{\label{fig:fig3}  Temperature-dependent (a) critical current density and (b) vortex creep parameter $S$ for Sample \#1 during pressurization. The insert shows the pressure dependence of $J_c$ and $S$ under different applied magnetic fields at 5 K. The data presented here is from the second pressurization cycle. }
\end{figure}

Figure~\ref{fig:fig3}(a) shows the temperature dependence of $J_c$ for Sample~\#1 at an applied field of 0.5~T during pressurization. As expected, $J_c$ decreases monotonically with increasing temperature. When comparing different pressures, the 0~GPa curve exhibits the lowest $J_c$, while all pressures above 0~GPa yield similar $J_c$ values, indicating only a weak pressure dependence, as summarized in the insert. We find that the sample volume decreases after pressurization. Because the geometric dimensions cannot be measured \textit{in situ} during the pressure cycle, a constant volume was assumed for all pressures in the Bean model calculation. As a result, the $J_c$ values shown for $P>0$~GPa are slightly underestimated. The exact dimensions used are provided in Supplementary Table~S1.

Many competing interactions dictate vortex dynamics, determining the rate of thermally-activated vortex motion (creep) and $J_c$. This includes vortex pinning by inhomogeneities in the free energy landscape (caused by defects or strain), vortex elasticity, vortex–vortex interactions, anisotropy, thermal fluctuations, and current‑driven forces \cite{Blatter1994, Feigelman1989, Yeshurun1996}. Pinning lowers the vortex core energy by $U_P$, while applied currents tilt the energy landscape \cite{PhysRev.131.2486}, reducing the effective barrier to a current‑dependent form $U(J)$. Considering the effects of current, temperature, and magnetic field, the activation barrier $U_{act}(T,H,J)$ follows an Arrhenius law $t = t_0 e^{U_{act}(J)/k_B T}$,  where $1/t_0$ is related to the attempt frequency estimated at $10^6 –10^{10}$ Hz \cite{Brandt1989b, Buchacek2019a, blatter_vortex_2003, Kwok2016}. Collective creep theory predicts that 
\begin{equation}\label{eq:Uact} U_{act}(J) = \frac{U_p}{\mu}\left[\left(\frac{J_c}{J}\right)^\mu - 1\right], 
\end{equation}
 where the glassy exponent $\mu$ characterizes the size and dimensionality of vortex bundles that hop during the creep process (e.g. single flux lines, small or large bundles of size less than or greater than the penetration depth) \cite{Yeshurun1996, Blatter1994}. This framework predicts logarithmic decay in the induced current density and corresponding magnetization 
\begin{equation}\label{eq:Mt} J(t) \propto M(t) \approx M_0\left[1+\frac{\mu k_B T}{U_p}\ln\left(\frac{t}{t_0}\right)\right]^{-1/\mu}. \end{equation} We can directly measure this decay in the magnetization, caused by vortex creep and termed magnetic relaxation, by repeatedly recording the magnetic moment $m(t) \propto M(t)$ over time. From this, we can extract the vortex creep parameter as follows:
\begin{align}\label{eq:ST} S \equiv \left| \frac{d \ln M}{d \ln t} \right| = \frac{k_B T}{U_0+\mu k_B T \ln (t/t_0)}. 
\end{align}

Figure~\ref{fig:fig3}(b) displays $S(T)$ during pressurization. Within the temperature range of $4-8$ K, $S$ systematically decreases with increasing pressure. This decreasing trend is more clearly illustrated in the insert, where both the 0.5 T and 1 T datasets exhibit a similar pressure dependence. The temperature dependence of $J_c$ and $S$ at $\mu_0 H = 1$~T is presented in Supplementary Fig.~S4.

\subsection*{Effect of pressure cycling on vortex pinning}

\begin{figure*}[!ht]
\centering
\includegraphics[width=1\linewidth]{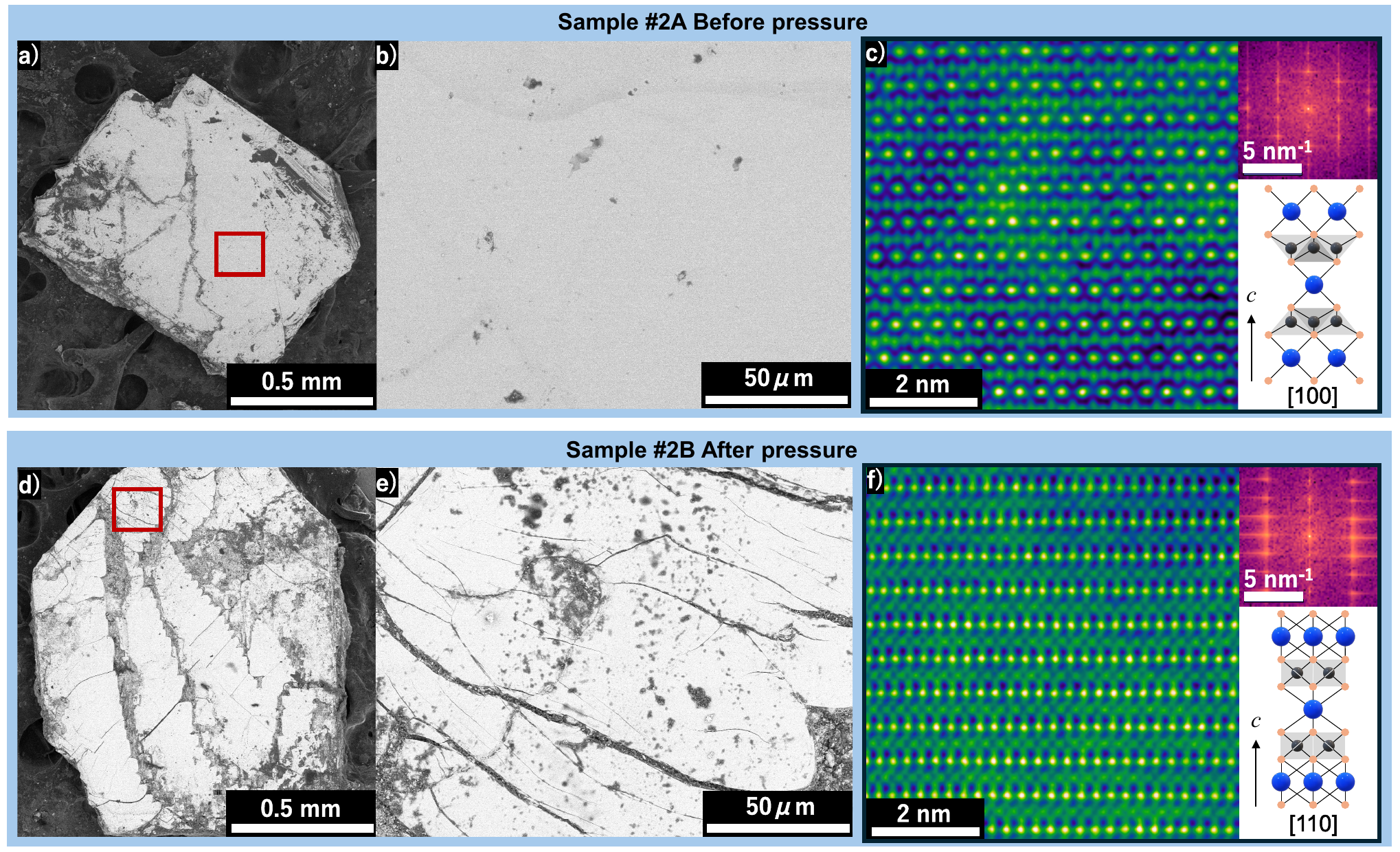}
\caption{\label{fig:fig4} Plan-view SEM images at two different magnifications are shown for (a,b) before and (d,e) after the pressure cycle. The red box indicates the location where the higher magnification is taken. Cross-sectional high-angle annular dark-field STEM images of Sample~\#2 for (c) before and (f) after the pressure cycle. A 3$\sigma$ Gaussian filter was applied to the images, with pixel sizes of (c) 15.6~pm/pixel and (f) 7.8~pm/pixel. The upper insets show the corresponding FFTs. The lower insets show the crystal structure of BaFe$_2$(As$_{1-x}$P$_{x}$)$_2$ viewed along the [100] direction in (c) and along the [110] direction in (f). The atom colors are Ba (blue), Fe (black), and As/P (orange). }
\end{figure*}

When performed at low temperature, the pressure-quench protocol has shown that pressure-induced structural modifications can be retained after pressure release at low temperatures, leading to significant enhancements in $T_c$ \cite{Deng2021,Deng2025}. Here, we study whether similar irreversible effects can occur on P-doped Ba122 crystals when pressurization and depressurization occur at room temperature. In Sample \#1, after two pressure cycles, with peak pressures of 0.60 and 1.08~GPa reached in cycles 1 and 2, respectively. We noticed that the pressurization process was irreversible, in that the $T_c$ and $J_c$ did not return to their original values after pressure release. To understand the origin of this irreversible pressure effect, we performed scanning electron microscopy (SEM), scanning transmission electron microscopy (STEM), and single-crystal X-ray diffraction (XRD) measurements on Sample~\#2. Sample~\#2 was cut into two pieces, which we designate as Samples \#2A and \#2B. Sample~\#2B was then pressurized from 0~GPa to 1.08~GPa and then released back to ambient pressure. 

We first used SEM to examine whether the pressure cycle altered the surface morphology. Compared with the surface before the pressure cycle (Fig.~\ref{fig:fig4}(a,b)), microcracks are observed after pressurization (Fig.~\ref{fig:fig4}(d,e)). These cracks likely originate from inhomogeneous pressure caused by the spacer compressing the sample in the uniaxial direction. For the largest observable cracks from the 65$\times$ image in Fig.~\ref{fig:fig4}(d), the crack density is approximately $21.1$~mm$^{-2}$, and the average crack length is $190 \pm 80~\mu$m. From the 650$\times$ image in Fig.~\ref{fig:fig4}(e), the crack density is approximately $1570$~mm$^{-2}$, and the average crack length is $24 \pm 11~\mu$m. The crack width was further estimated using the 3500$\times$ and 12000$\times$ images in Supplementary Fig.~S7(d-f). The widest cracks are about $2~\mu$m, and the narrower ones range from 100 to 500~nm.

We next used STEM to examine whether the pressure cycle altered the nanoscale defect landscape, as shown in Fig.~\ref{fig:fig4}(c,f), and observed no changes. However, it is important to note that the STEM measurements probe only a limited region of the sample. Therefore, they may not reflect the full sample defect landscape. We then measured the interatomic spacings using the STEM images and their corresponding Fast Fourier Transforms (FFTs) (upper insets of Fig.~\ref{fig:fig4}(c,f)). The extracted \textit{c}-lattice parameters before and after the pressure cycle are $1.274 \pm 0.002$~nm and $1.274 \pm 0.001$~nm, respectively, indicating no measurable change within experimental uncertainty. The \textit{c}-plane single-crystal XRD results in Supplementary Fig.~S8 also show no change in \textit{c}-lattice parameters. Likewise, the \textit{a/b}-lattice parameters are $0.3894 \pm 0.0013$~nm and $0.3879 \pm 0.0016$~nm for before and after the pressure cycle, respectively, again showing no measurable difference. The \textit{a}- and \textit{b}-lattice parameters are equal, and all measured values agree with the previous report~\cite{Allred2014}. Larger field-of-view STEM images and convergent-beam electron diffraction results are provided in Supplementary Fig.~S7(b,c,g).

\begin{figure*}[!ht]
\centering
\includegraphics[width=1\linewidth]{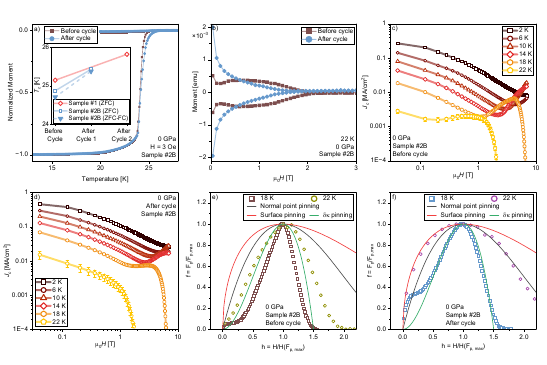}
\caption{\label{fig:fig5} (a) Normalized magnetic moment as a function of temperature for Sample \#2B before and after the first pressurization cycle and release back to ambient pressure. The inset shows the effect of the pressure cycle on $T_c$ for both Sample \#1 and Sample \#2B, where $T_c$ values were extracted using two approaches: ZFC-only, and from the difference between the ZFC and FC curves. For Sample \#1, the peak pressure achieved in cycles 1 and 2 is 0.60 and 1.08 GPa, respectively. For Sample \#2B, the peak pressure achieved after the cycle is 1.08 GPa. (b) Effect of the pressure cycle on the magnetic hysteresis loop of Sample \#2B at 22 K. Field-dependent $J_c$ for Sample \#2B measured (c) before and (d) after the pressure cycle at different temperatures. Normalized pinning force $f=F_p/F_{p,max}$ against reduced field $h=H/H(F_{p, max})$ at 18 and 22 K for (e) before and (f) after the pressure cycle. }
\end{figure*}

We then measured $T_c$, $J_c$, and $S$ in Sample~\#2B before pressurization and after pressurization. Figure~\ref{fig:fig5}(a) compares the temperature-dependent moment in each case, revealing a noticeable increase in  $T^{onset}_c$ for both samples after the pressure cycle, plotted in the inset. Magnetic hysteresis loops were measured at temperatures of $2–22$ K before and after the pressure cycles. The 22 K results are shown in Fig.~\ref{fig:fig5}(b), and the remaining curves are provided in Supplementary Fig.~S2. A clear second magnetization peak (SMP) is observed between $2–18$ K for both before and after pressure cycling. However, a striking change occurs at 22 K, where the SMP fully disappears after the pressure cycle. In the Ba122 family, such as P-, Ni-, K-, and Co-doped compounds, the SMP has been associated with a crossover from collective (elastic) to plastic vortex creep~\cite{Sundar2017,Sundar2019,miu_smp_2020, liu_doping_2018,liu_thickness_2024}, often accompanied by a rhombic-to-square vortex lattice transition~\cite{rosenstein_peak_2005,sundar_study_2017}.

To investigate the disappearance of the SMP, we analyzed the vortex pinning mechanism using the pinning force $F_p = \mu_0 H J_c$~\cite{Kramer1973,Hughes1974,Yamasaki1993,Koblischka1998,Yang2008,Yamamoto2009,Fang2011,Luo2025}. We first calculated the field-dependent critical current density $J_c(B)$ using the Bean critical state model Eq.(\ref{eq:Bean}) ~\cite{Gyorgy1989,Talantsev2024BeanModel}, shown in Fig.~\ref{fig:fig5}(c–d). These results clearly demonstrate that the SMP shifts and flattens after the pressure cycle and ultimately disappears at 22 K. Meanwhile, the overall $J_c$ increases, indicating enhanced vortex pinning. Next, we plot the normalized pinning force $f = F_p/F_{p,\mathrm{max}}$ against the reduced field $h = H/H(F_{p,\mathrm{max}})$ as shown in Fig.~\ref{fig:fig5}(e,f). Here, $F_{p,\mathrm{max}}$ is the maximum pinning force, and $H(F_{p,\mathrm{max}})$ is the field at which this maximum occurs. According to the Dew-Hughes model, the normalized pinning force for three common pinning cases results in the following function forms \cite{Hughes1974, Luo2025}:
\begin{align}
    \delta\kappa \text{ point pinning:} \quad & f = 3 h^2 \left(1 - \frac{2h}{3}\right) \label{eq:dk}\\
    \text{Normal point pinning:} \quad & f = \frac{9}{4} h \left(1 - \frac{h}{3}\right)^2 \label{eq:normpin}\\
    \text{Surface pinning:} \quad & f = \frac{25}{16} h^{1/2} \left(1 - \frac{h}{5}\right)^2 \label{eq:surfpin}
\end{align}

The \(\delta\kappa\) point pinning mechanism arises when pinning sites locally modify the Ginzburg--Landau parameter,$\kappa = \lambda / \xi$, relative to the bulk value, producing a spatial variation in \(\kappa\) that generates a vortex pinning potential. Here, $\lambda$ is the penetration depth and $\xi$ is the coherence length. Normal point pinning is caused by defects such as non-superconducting inclusions or point defects and can be associated with $\delta T_c$ pinning ~\cite{Fang2011}. As the temperature approaches $T_c$, variations in the local transition temperature become increasingly important, creating regions of reduced condensation energy. Surface pinning originates from extended two-dimensional defects, including grain and twin boundaries, stacking faults, and dislocation arrays such as sub-grains. As well as plate-like precipitates, and even the surface of the superconductor.

Figure~\ref{fig:fig5}(e–f) shows the $f(h)$ curves before and after the pressure cycle, plotted together with the three Dew-Hughes pinning models described by Eqs. (\ref{eq:dk}-\ref{eq:surfpin}). For temperatures $T \leq 14$~K, the pinning force $F_p(H)$ does not exhibit a well-defined maximum, and therefore a meaningful $F_{p,\mathrm{max}}$ cannot be extracted. For this reason, only the 18 and 22~K data are shown. Before the pressure cycle, both the 18 and 22~K curves closely follow the $\delta\kappa$ pinning model, consistent with previous studies on the P-doped Ba122 system \cite{Fang2011}. This pinning behavior has been suggested to originate from nanoscale inhomogeneity in the dopant distribution, which leads to fluctuations in the electron mean free path \cite{Beek2010, Chong2010, Fang2011, Shabbir_MgB2_2015}.

After the pressure cycle, the 18~K data remain close to the $\delta\kappa$ model. However, at 22~K the pinning mechanism changes significantly. For $h > 1$, the curve follows the normal point ($\delta T_c$) pinning model. For $h < 1$, the 22~K data instead align more closely with the surface pinning model. This is consistent with the SEM results, since the pressure cycle alters the surface morphology and produces a high density of microcracks. These microcracks may act as extended surface defects that locally enhance the screening current density near the crack edges, thereby facilitating vortex entry \cite{aladyshkin2001}. We also note that at 18 K both before and after the pressure cycle, the $f$ vs $h$ data exhibit a plateau near $h \approx 0.3$, a feature also reported in Ref.~\cite{Fang2011} and attributed to strong point pinning. Therefore, the disappearance of the SMP at 22 K may be attributed to a pressure-induced modification of the vortex lattice that changes the pinning mechanism from primarily $\delta\kappa$ pinning to a mixed combination of $\delta T_c$ and surface pinning.

\begin{figure*}[!ht]
\centering
\includegraphics[width=0.8\linewidth]{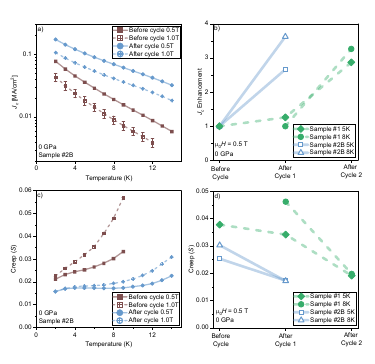}
\caption{\label{fig:fig6} Temperature-dependent (a) $J_c$ and (c) $S$ for Sample \#2B measured before and after the pressure cycle under applied fields of $\mu_0H = 0.5$ and 1 T. (b) Enhancement of $J_c$ and (d) the effect of the pressure cycle on $S$ for both Sample \#1 and Sample \#2B at 5 K and 8 K under an applied field of $\mu_0H = 0.5$ T. Both samples exhibit more than a threefold increase in $J_c$ at 8 K and 0.5 T after pressure cycling.}
\end{figure*}

We next calculated the temperature dependence of the critical current density $J_c(T)$ using Eq.(\ref{eq:Bean}), and the vortex creep parameter $S(T)$ extracted from magnetic relaxation measurements. As shown in Fig.~\ref{fig:fig6}(a–b), pressure cycling results in a dramatic enhancement of vortex pinning, the critical current density increases by more than a factor of three at 8 K and $\mu_0H = 0.5$~T, and remains substantially higher across the full temperature window from 2–14 K. For Sample~\#1, the same post-cycle sample dimensions were used in the $J_c$ calculation for all pressure cycles, since dimensional changes could not be directly measured during pressurization. Therefore, any pressure-induced volume reduction is not explicitly included, and the reported $J_c$ enhancement should be regarded as a conservative estimate. In contrast, for Sample~\#2, the sample dimensions were directly measured both before and after the pressure cycle, so the reported $J_c$ values and enhancements are quantitative within experimental uncertainty. Concomitantly, the creep parameter $S(T)$ is strongly suppressed, as shown in Fig.~\ref{fig:fig6}(c–d). For Sample~\#1, at $\mu_0H = 0.5$~T and 8~K, $S$ drops from $\sim0.046$ to $\sim0.02$. In Sample~\#2B, at 1~T and 8~K, $S$ decreases from $\sim0.048$ to $\sim0.019$, and even at 0.5~T, $S$ drops from $\sim0.03$ to $\sim0.017$. Sample~\#2B shows a smaller reduction in creep at 0.5~T compared to Sample~\#1. We attributed this to the pressure environment during cycling. Since the Teflon cell is not reusable, variations in sample dimensions and loading configurations between experiments may lead to different degrees of pressure inhomogeneity. This may affect the microcrack density, consequently influencing the vortex dynamics.

\begin{figure*}[!ht]
\centering
\includegraphics[width=1\linewidth]{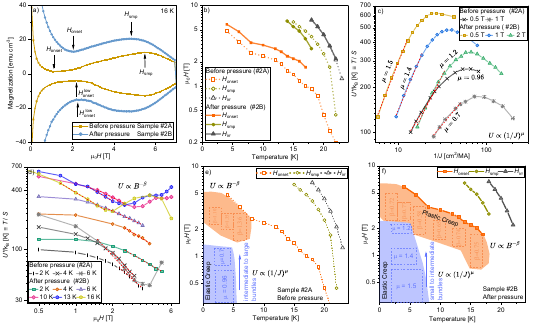}
\caption{\label{fig:fig7} (a) Magnetic hysteresis loop of Sample~\#2 at 16~K, with the second magnetization peak field $H_{\mathrm{smp}}$, the upper ($H_{\mathrm{onset}}$) and lower ($H^{\mathrm{low}}_{\mathrm{onset}}$) branch onset indicated by black arrows. (b) Effect of the pressure cycle on $H_{\mathrm{onset}}$, $H_{\mathrm{smp}}$, and the irreversibility field $H_{\mathrm{irr}}$ of Sample~\#2 at different temperatures. (c) Energy scale $U^*/k_B = T/S$ versus $1/J$ for Sample~\#2 before and after the pressure cycle on a logarithmic scale. Linear fits (red dashed lines) to the $U^*(1/J)$ data were used to extract $\mu$, based on $U^* \propto (1/J)^{\mu}$. (d) Energy scale $U^*/k_B = T/S$ versus $B$ for Sample~\#2 before and after the pressure cycle on a logarithmic scale. Linear fits (red dashed lines) to the $U^*(B)$ data were used to extract $\beta$, based on $U^* \propto B^{-\beta}$. Vortex phase diagrams of Sample~\#2 (e) before and (f) after the pressure cycle.}
\end{figure*}

Lastly, we examined the effect of the pressure cycle on vortex dynamics by constructing the vortex phase diagram for Sample~\#2. The field at which the second magnetization peak appears, $H_{\mathrm{smp}}$, and its onset field, $H_{\mathrm{onset}}$, were determined from the upper branches of the hysteresis loops, as illustrated in Fig.~\ref{fig:fig7}(a). After the pressure cycle, both characteristic fields shift noticeably, as marked by the black arrow. We also observe that the pre-pressure loop shows strong asymmetry, evidenced by the difference in $H_{\mathrm{onset}}$ between the upper and lower branches, while the post-pressure loop is more symmetric. Asymmetry in the hysteresis loop is commonly associated with different vortex entry and exit processes in the superconductor due to geometrical and surface barriers\cite{Zeldov1994,Benkraouda1996,Chen1993, Burlachkov1993}. To characterize this systematically, the difference between the SMP onset fields on the upper and lower branches, $\Delta H_{\mathrm{onset}}$, is evaluated in Supplementary Fig.~S9. The average $\Delta \bar{H}_{\mathrm{onset}}$ over 4–17 K (SMP disappears above 18 K) decreases from 1.19 T before the pressure cycle to 0.58 T after the pressure cycle, indicating a clear increase in loop symmetry. We attribute this enhanced symmetry suppression of the Bean–Livingston surface barrier (by the pressure-induced cracks), which facilitates vortex entry, thereby reducing the asymmetry between flux entry and exit\cite{aladyshkin2001}.

We then plotted the temperature-dependent $H_{\mathrm{onset}}$, $H_{\mathrm{smp}}$, and irreversibility field $H_{\mathrm{irr}}$ before and after the pressure cycle in Fig.~\ref{fig:fig7}(b). Here, $H_{\mathrm{irr}}$ is extracted from $J_c(B)$ and defined when $J_c < 2\times10^{-5}$~MA~cm$^{-2}$. We find that the $H_{\mathrm{onset}}$ line shifts to a higher field, whereas the $H_{\mathrm{smp}}$ line shifts to a lower field after the pressure cycle. Additionally, the SMP disappears in the post-pressure sample above 18~K. Since the SMP in the Ba122 family has been associated with a crossover from collective (elastic) to plastic vortex creep\cite{Sundar2017,Sundar2019,liu_doping_2018,miu_smp_2020,liu_thickness_2024,rosenstein_peak_2005,sundar_study_2017}, these opposite shifts suggest that the pressure cycle suppresses the crossover regime and reduces the degree of vortex-lattice disorder. In contrast, $H_{\mathrm{irr}}$ remains nearly unchanged after the pressure cycle, showing that the pressure-induced modifications do not affect the transition to a vortex-liquid state.

To extract the elastic creep region and the creep bundles size using $\mu$, it is common practice\cite{Zhou2016a, Sun_2015, Sun2015d,Haberkorn2011b,Miu2013,Sundar2017} to define an experimentally accessible auxiliary energy scale $U^* \equiv U_0+\mu k_BT\ln(t/t_0)= k_BT/S$ (see Eq. (\ref{eq:ST})), such that combining Eqs. (\ref{eq:Uact}) and $t = t_0 e^{U_{act}(J)/k_B T}$ is consistent with $U^* =U_0(J_{c0}/J)^\mu$. Figure~\ref{fig:fig7}(c) plots $U^*$ vs $1/J$ on a logarithmic scale for both before and after sample pressurize cycling. We obtain $\mu$ from the slope in the linear region. Before the pressure cycle, $\mu = 0.96$ at 0.5~T and 0.7 at 1~T, consistent with a transition from intermediate-bundle to large-bundle creep with increasing magnetic field\cite{Blatter1994,Abulafia1996}. After the pressure cycle, $\mu = 1.5$ at 0.5~T and 1.2 at 2~T, consistent with a transition from small-bundle to intermediate-bundle creep\cite{Blatter1994,Abulafia1996}.

To further examine the high-field creep regime, we plot $U^*/k_B$ as a function of $B$ on a logarithmic scale and extract the exponent $\beta$ from the linear region using $U^* \propto B^{-\beta}$, as shown in Fig.~\ref{fig:fig7}(d). Theory on vortex lattice dislocation-mediated plastic creep predicts the activation energy $U_{pl}\propto B^{-1/2, -3/4}$ \cite{Abulafia1996, Burlachkov2022}. For Sample~\#2 after the pressure cycle, $\beta$ falls in the range $0.58-0.77$, which is consistent with dislocation-mediated plastic creep. In contrast, before the pressure cycle, $\beta$ is larger, ranging from 1.1 to 1.36. This stronger field dependence deviates from the simple dislocation-mediated plastic creep prediction and may indicate additional mechanisms, such as surface barriers.

Finally, vortex phase diagrams were constructed for the sample before and after the pressure cycle using the elastic- and plastic creep analyses shown in Fig.~\ref{fig:fig7}(e,f). Here, the elastic creep region is clearly expanded after the pressure cycle. Additionally, the post-pressure cycle phase diagram indicates that $H_{\mathrm{onset}}$ marks the upper boundary of the plastic creep. This combination of a threefold enhancement in $J_c$, a more than 40\% reduction in the creep parameter, and an expanded elastic creep region demonstrates that pressure cycling can simultaneously strengthen vortex pinning and improve its stability, with these enhancements fully retained after pressure release.

\section*{Conclusions}

In summary, we have shown that near-hydrostatic pressures applied using a compact, off-the-shelf pressure cell can produce microcracks and substantially and irreversibly enhance vortex pinning in overdoped BaFe$_2$(As$_{1-x}$P$_x$)$_2$ single crystals. While pressures up to 1.08 GPa produce only a small suppression of $T_c$, they lead to a more than threefold increase in $J_c$ and over a 40\% reduction in the creep parameter $S$ at 8 K and 0.5 T, with these improvements retained after pressure release at room temperature. Pressure cycling also modifies the pinning landscape, removing the second magnetization peak above 18 K and providing evidence for a transition from predominantly $\delta\kappa$ pinning to a mixed $\delta T_c$ and surface pinning, while also expanding the elastic-creep region. These results establish pressure cycling as a simple route to stabilize enhanced vortex pinning without changing composition or introducing artificial pinning centers.

Unlike the low-temperature pressure quenching protocol, the present approach preserves the enhancement even after pressure is released at room temperature, making it far more practical for real‑world applications. Moreover, applying this protocol to other iron-based superconductors will help test the generality of irreversible pressure-induced effects and may identify compositions that yield even larger improvements in $J_c$ and creep suppression under practical operating conditions.

\section*{Methods}\label{sec:Methods}
\subsection*{Crystal Growth}
Single crystals of BaFe$_2$(As$_{1-x}$P$_x$)$_2$ were grown using a two-step self-flux method. Binary precursors Ba$_2$As$_3$, Ba$_2$P$_3$, FeAs, and FeP were first synthesized from high-purity elements. These precursors were then mixed with Ba pieces in controlled ratios, sealed in evacuated quartz tubes, and heated to high temperatures. Single crystals were obtained upon slow cooling. The detailed growth sequence and mixture ratios are described in Ref. \cite{Nakajima2012, Zhang2019}. Energy-dispersive X-ray spectroscopy (EDS) was used to verify the doping level.

\subsection*{STEM imaging}
Atomic resolution STEM imaging was performed using a Nion UltraSTEM 100 aberration corrected microscope operating at 60 kV acceleration voltage and a convergence semi-angle of 33mrad. A high annular angle dark field (HAADF) detector was used to obtain Z-contrast images with a pixel size of about 8 pm/pixel and a dwell time of 4 $\mu$s. A probe current of 10 pA was used to acquire the images. The collection angle of the HAADF detector was set to between 80 to 200 mrad. In order to correct sample drift, 20 HAADF frames were acquired, aligned, and integrated to obtain the final image with high signal to noise ratio. The convergent beam electron diffraction pattern was collected using a 3 mrad convergence semi-angle and recorded using a Nion 2020 Ronchigram camera, with a Hamamatsu ORCA ultra-low noise scientific CMOS sensor with a 20 ms exposure time.

\subsection*{Magnetometry Measurements}
Magnetization measurements were performed using a Quantum Design MPMS3 SQUID magnetometer. Sample \#1 was measured inside the Quantum Design high pressure cell module, whereas Sample \#2 was pressurized using the pressure cell but measured in the standard MPMS3 brass sample holder. The field- and temperature-dependent background signal of the brass sample holder is included in Supplementary Fig.~S5. In all measurements, the magnetic field was applied perpendicular to the film plane (parallel to the $c$-axis, $H \parallel c$). For magnetic hysteresis and relaxation measurements $m(H)$, each moment scan was performed over a 25~mm scan length in 4~s, while for moment versus temperature measurements $m(T)$ used to extract $T_c$, a 10 mm scan length collected in 1 s was used.

To determine $T_c$, the magnetic moment $m(T)$ was measured under $\mu_0H = 0.2-1$ mT while sweeping temperature at approximately 0.05 K min$^{-1}$ for measurements conducted in the pressure cell and 1.5 K min$^{-1}$ for measurements using the brass sample holder. Magnetic hysteresis loops $m(H)$ were acquired by stabilizing the magnetic moment at each field after sweeping at a rate of 100~Oe~s$^{-1}$.

Magnetic relaxation (vortex creep) was obtained following conventional protocols \cite{Yeshurun1996}, in which the moment $m(t,H_1,T_1)$ was recorded every 10~s for approximately one hour after establishing a critical state. The critical state was established by sweeping the field by $\Delta H = 1-2$~T~$> 4H^*$, where $H^*$ is the minimum flux-penetration field, and then holding the field at $H_1$ and temperature at $T_1$. The preparation of the critical state was confirmed by comparing the initial $m(t)$ with the corresponding hysteresis loop $m(H)$. After subtracting the background contribution from the sample mount and correcting for the time offset between field application and the first recorded point, the creep parameter was extracted using $S = -d\ln m/d\ln t$ from a linear fit to $\ln m$ versus $\ln t$. The detailed procedure for processing the relaxation data and additional $S(T)$ and $S(H)$ plots are provided in Supplementary Fig.~S3. Lastly, a comparison of the magnetometry measurements on Samples~\#2A and \#2B before pressure cycling, showing their consistency, is included in Supplementary Fig.~S10.

\section*{Data availability}

The data supporting the findings of this study are available on Mendeley Data (DOI: 10.17632/j5x7z8nzhy.1) as a zip file. This includes Python code used to process the data and Origin files (.opju) that contain data spreadsheets for all the samples and figures used in this paper, which can be opened using Origin Viewer, a free application that permits viewing and copying of data contained in Origin project files. The code used for pressure cell background subtraction is adapted from the open source SquidLab program \cite{SQUIDLAB}.

\section*{Acknowledgments}
This material is based upon work supported by the National Science Foundation under the University of Washington Materials Research Science and Engineering Center under grant DMR-2308979 (J.L., A.L., D.S., G.R., Z.L., J.I., J.C., S.E.). 

\vspace{0.3 cm}

\section*{Author Contributions}
S.E. conceived and designed the experiment.
D.S., G.R., and J.I. acquired the STEM and convergent-beam electron diffraction images.
Z.L. and J.C. grew the P-doped Ba122 crystals.
J.L. performed EDS, XRD, and SEM on crystals.
J.L. and A.L. performed magnetization studies and data analysis.
J.L. developed the background subtraction procedure and wrote the associated code.
S.E. and J.L. thoroughly reviewed the data analysis.
J.L. and S.E. wrote the manuscript.
All authors commented on the manuscript.

\pagebreak
\let\clearpage\relax 
\onecolumngrid
\section*{Supplemental Materials}\label{sec:smaterials}
\setcounter{figure}{0}

\makeatletter 
\renewcommand{\thefigure}{S\@arabic\c@figure}
\makeatother

\setcounter{figure}{0}
\makeatletter 
\renewcommand{\thefigure}{S\@arabic\c@figure}
\makeatother

\setcounter{table}{0}
\makeatletter 
\renewcommand{\thetable}{S\@arabic\c@table}
\makeatother

\newcommand{\vect}[1]{\mathbf{#1}}
\DeclareSIUnit\oersted{Oe}

\maketitle

\section*{1. Pressure calibration}

\begin{figure*}[h!]
\centering
\includegraphics[width=1\linewidth]{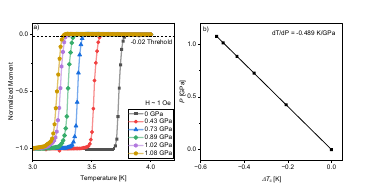}
\caption{\label{fig:figS1} (a) Normalized moment against temperature for the tin manometer at different pressures. (b) Pressure against $\Delta T_c=T_c -T_c(0$ GPa). Using the linear pressure dependent on tin $dT_c/dP=-0.489$ K/GPa, pressure can be calculated using $P=\Delta T_c/-0.489$ K/GPa. }
\end{figure*}

A moment against temperature measurement was performed on the tin manometer between 3–4 K under zero field cooled conditions to determine its $T_c$. The $T_c$ is defined as the temperature at which the normalized magnetic moment drops below the threshold value of $-0.02$, as shown in Fig.~\ref{fig:figS1}(a). The applied pressure is then calculated using $P = \Delta T_c/(-0.489)$, where $\Delta T_c = T_c - T_c(\text{0 GPa})$, as shown in Fig.~\ref{fig:figS1}(b).

\section*{2. Magnetic hysteresis}

\begin{figure*}[h!]
\centering
\includegraphics[width=1\linewidth]{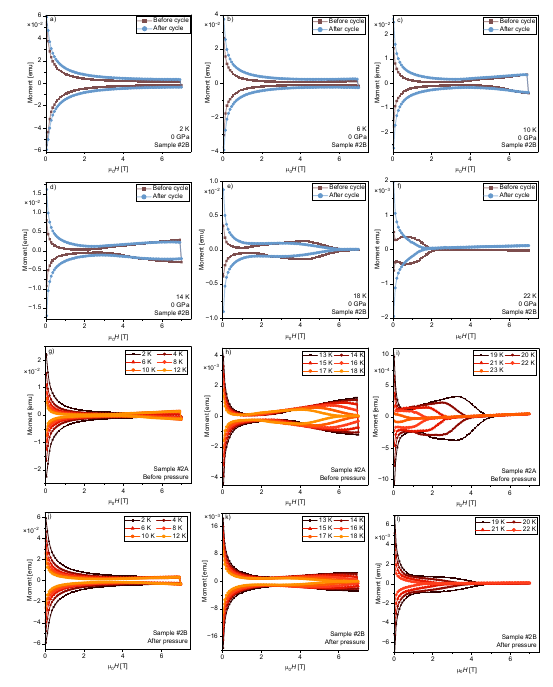}
\caption{\label{fig:figS2} Effect of the pressure cycle on the magnetic hysteresis loop of Sample \#2B at (a) 2 K, (b) 6 K,(c) 10 K,(d) 14 K,(e) 18 K,(f) 22 K. Additional hysteresis loops for Sample (g-i) \#2A before pressure, (j-l) Sample \#2B after pressure. }
\end{figure*}

Magnetic hysteresis loops measured before and after the pressure cycle for sample~\#2B between 2 and 22 K are shown in Fig.~\ref{fig:figS2}(a–f). For temperatures between 2 and 10 K, the pressure cycle increases the magnetic moment across all applied fields. At intermediate temperatures $14-18$ K, before the second magnetization peak (SMP), the pressure cycle enhances the moment, whereas above the SMP, the moment is reduced after pressure cycling. We also observed that the SMP shifts to lower fields, as seen in Fig.~\ref{fig:figS2}(d, e). At 22 K, the SMP disappears entirely, as shown in Fig.~\ref{fig:figS2}(f). Additional loops for Sample \#2A and \#2B are shown in Fig.~\ref{fig:figS2}(g-l).

\section*{3. Magnetic relaxation}

\begin{figure*}[h!]
\centering
\includegraphics[width=1\linewidth]{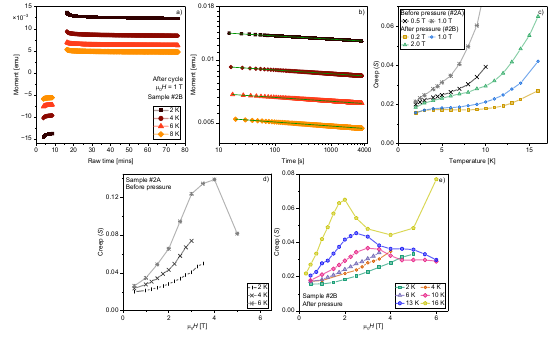}
\caption{\label{fig:figS3} (a) Example raw magnetic relaxation curves $m(t)$ for sample~\#2B after the first pressure cycle, measured at $\mu_0 H = 1$~T for various temperatures. (b) The same data after correcting for the system delay time $t_d$. The resulting plots show $\ln m$ versus $\ln (t + t_d)$, where the slope of the linear fit defines the relaxation rate $S$. Additional (c) temperature- and (d, e) Field-dependent vortex creep parameter of Sample \#2 for before and after pressure cycle.}
\end{figure*}

Figure~\ref{fig:figS3}(a) shows the example raw magnetic relaxation data for sample~\#2B after the first pressure cycle, measured under an applied field of $\mu_0 H = 1$~T. Relaxation was recorded for 5 minutes on the lower branch and for 60 minutes on the upper branch of the hysteresis loop, in both cases after preparing the sample in the critical state. Because the MPMS3 acquires the first data point after a short delay, during which relaxation has already begun. Therefore, we introduce a system delay time $t_d$ when extracting the vortex creep rate. The creep parameter is defined as $S = -d \ln m/d \ln (t + t_d)$,
where $t_d$ is treated as a fitting parameter chosen to maximize the linear correlation. The fully processed moment data used to obtain $S$ are shown in Fig.~\ref{fig:figS3}(b).

\begin{figure*}[h!]
\centering
\includegraphics[width=1\linewidth]{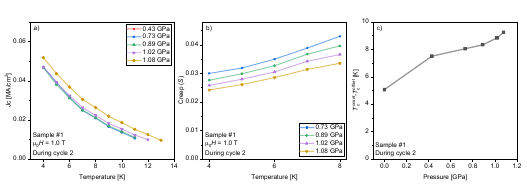}
\caption{\label{fig:figS4} 
(a) Critical current density $J_c$ as a function of temperature for sample~\#1 during pressure cycle~2 at $\mu_0 H = 1$~T. Here, $J_c$ is calculated using Bean's critical state model from the first data point of the magnetic relaxation measurement. (b) Vortex creep parameter $S$ as a function of temperature for sample~\#1 during pressure cycle~2 at $\mu_0 H = 1$~T. (c) Pressure dependent of transition width ($T^{onset}_c - T^{offset}_c$) for sample~\#1.}
\end{figure*}

Figure~\ref{fig:figS4}(a) presents additional results of the temperature dependence of $J_c$ at $\mu_0 H = 1$~T for sample~\#1, calculated using the Bean model from the initial magnetization value in the relaxation measurement. Figure~\ref{fig:figS4}(b) shows the corresponding temperature dependence of the vortex creep parameter $S$ for sample~\#1. Lastly, the transition width ($T^{onset}_c - T^{offset}_c$) during pressurization for sample~\#1 is plotted in Fig.~\ref{fig:figS4}(c). We observe a sudden increase in the transition width once pressure is applied, which corresponds to the relatively broadened transition shown in Fig.~2(a) of the main text. Both effects could be linked to pressure inhomogeneity developing from the filler material inside the pressure cell.

\section*{4. Magnetic background of the MPMS 3 brass sample holder}

\begin{figure*}[h!]
\centering
\includegraphics[width=0.7\linewidth]{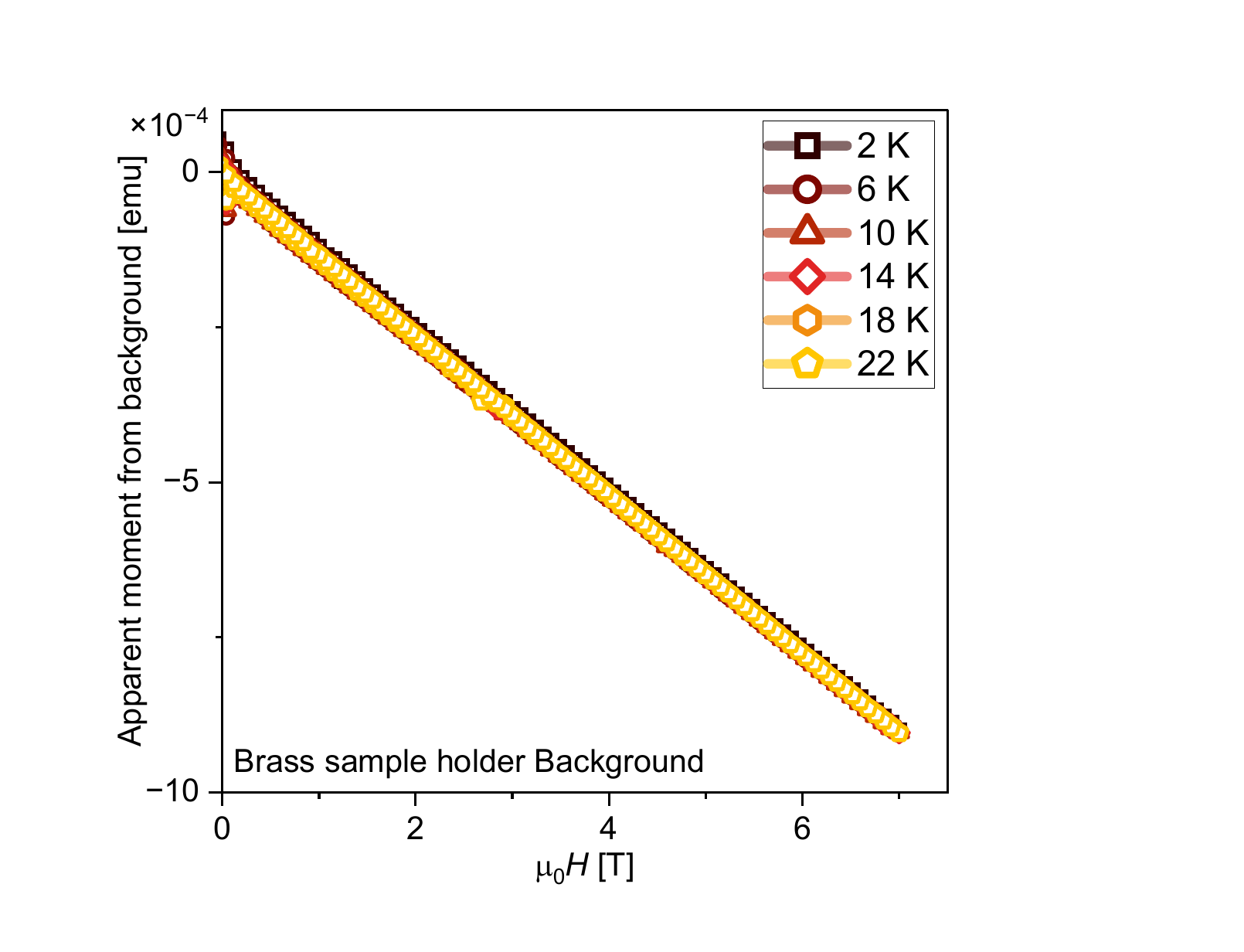}
\caption{\label{fig:figS5} Field-dependent background moment of the brass sample holder measured at different temperatures. The “apparent moment” refers to the moment parameter returned by the dipole fit applied to a non-dipolar background response.}
\end{figure*}

Sample~\#1 was measured using the pressure cell module, whereas Sample \#2B was pressurized using the pressure cell but measured in the standard MPMS3 brass sample holder. It is therefore important to characterize the background signal of the brass sample holder. Here, we measured the field-dependent background between 2 and 22 K in increments of 4 K, as shown in Figure~\ref{fig:figS5}. The brass sample holder exhibits a much weaker background, showing a linear field dependence and reaching only about $9\times10^{-4}$~emu at 7~T. In addition, it shows no measurable temperature dependence between $2$ and $22$~K. Although the background moment is small, to maintain consistency across all measurements in this study, voltage–position background subtraction was also performed for all data collected using the brass sample holder. Since the brass sample holder shows no measurable temperature dependence in the background signal, a single background field was used for each temperature interval. Specifically, the 4 K background was applied to data between $2$--$4$ K; the 6 K background to $5$--$8$ K; the 10 K background to $9$--$12$ K; the 14 K background to $13$--$16$ K; the 18 K background to $17$--$20$ K; and the 22 K background to $21$--$22$ K.

\section*{5. Magnetic background of the 1.3 GPa Model high pressure cell}

When collecting magnetic relaxation data using the pressure cell, apart from investigating the temperature- and field-dependent background as shown in the main text, it is also important to determine whether the background is time-dependent. To evaluate this, a magnetic relaxation sequence identical to that used for the sample measurements was performed on the empty pressure cell at 4, 5, 6, 8, 10, 12, 14, and 16 K under applied fields of 0.5 and 1 T. At each temperature and field, relaxation data were collected for 3 minutes on both the upper and lower branches of the hysteresis loop, as shown in Figure~S6. The pressure cell background exhibits no noticeable time dependence, except for the first data point (before 10 s), with a moment drop of $2\times10^{-5}$ emu, which is likely associated with eddy-current effects in the cell. 

As discussed in the main text, the pressure cell background shows strong temperature and field dependence. Therefore, background subtraction for the relaxation measurements was performed using background data obtained at the closest matching temperature and field. For measurements at 4, 5, 6, 8, 10, 12, 14, and 16 K, the corresponding background data were directly used. For intermediate temperatures (7, 9, 11, 13, and 15 K), the voltage–position background was obtained by linear interpolation between the nearest measured temperatures (e.g., 6 and 8 K for 7 K). For the 2–3 K data, the 4 K background was used.

When the background signal is comparable to the sample signal, background subtraction is required, as presented in this work. However, we cannot reliably extract the sample signal and perform a dipole fit once the moment from the background surpasses that produced by the sample. To quantify the robustness of the dipole fitting procedure, we define a fit quality factor $Q$ as

\begin{align}
Q = 1 - \frac{\max \left( \left| V - V_{\mathrm{fit}} \right| \right)}
{\max \left( \left| V \right| \right)}
\label{eq:fit_quality}
\end{align}

where $V$ is the measured voltage and $V_{\mathrm{fit}}$ is the fitted dipole voltage. The extracted magnetic moment is considered robust when $Q > 0.8$.

\begin{figure*}[h!]
\centering
\includegraphics[width=1\linewidth]{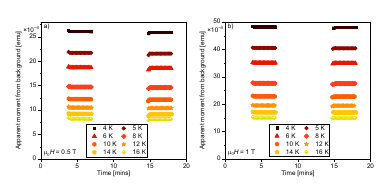}
\caption{\label{fig:figS6} Time-dependent background moment of the brass sample holder measured at different temperatures under (a) $\mu_0H = 0.5$ T and (b) $\mu_0H = 1$ T.
The “apparent moment” refers to the moment parameter returned by the dipole fit applied to a non-dipolar background response.}
\end{figure*}

\begin{figure*}[h!]
\centering
\includegraphics[width=1\linewidth]{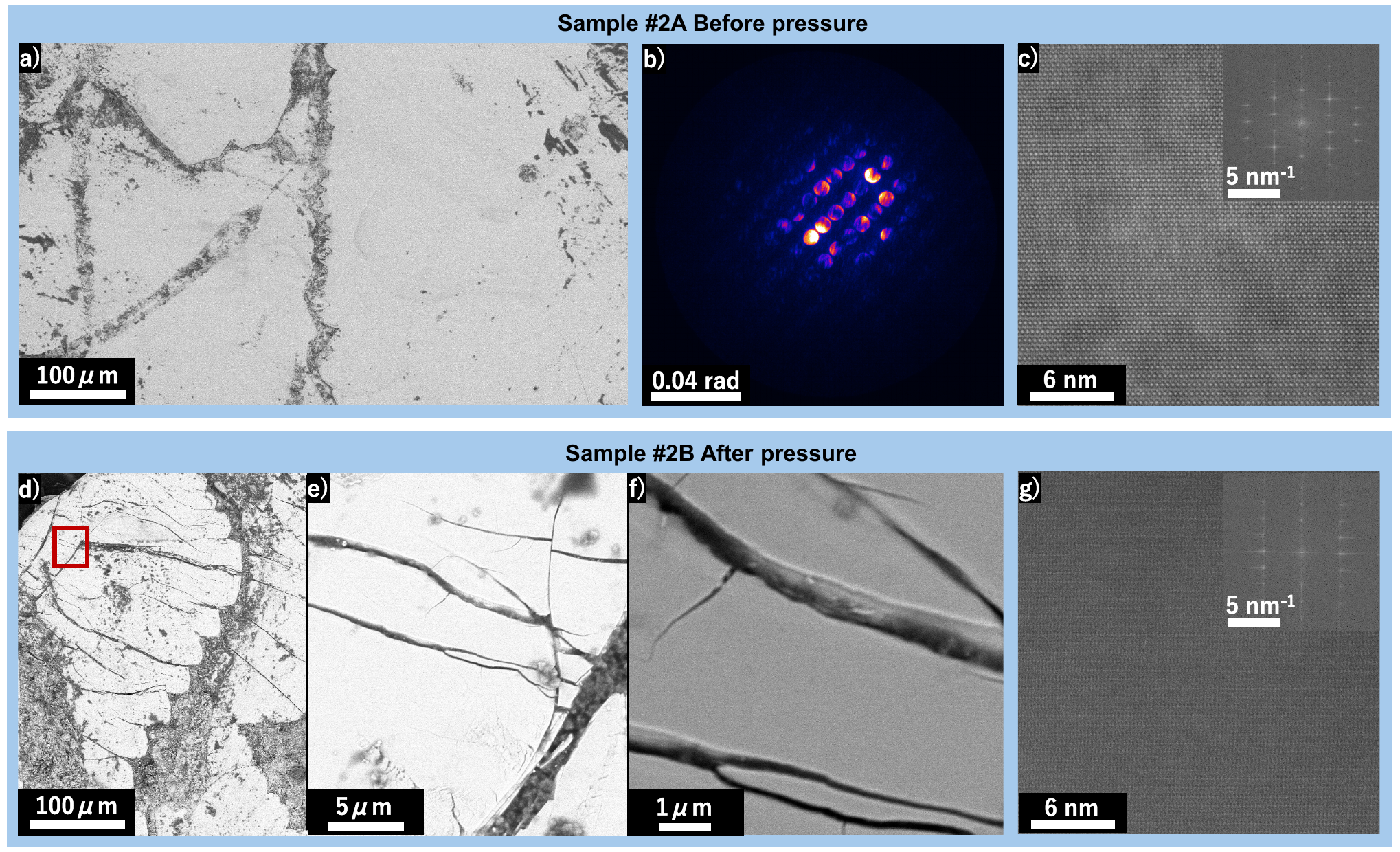}
\caption{\label{fig:figS7} Plan-view surface morphology SEM images at different magnifications are shown for (a) before and (d-f) after the pressure cycle. The red box indicates the location where the higher magnification is taken. (b) Convergent beam electron diffraction of Sample~\#2A before pressure cycle. Cross-sectional high-angle annular dark-field STEM images of Sample~\#2 for (c) before and (g) after the pressure cycle. The insets show the corresponding fast Fourier transforms (FFTs).  }
\end{figure*}

\section*{6. Single crystal X-ray diffraction}

\begin{figure*}[h!]
\centering
\includegraphics[width=0.6\linewidth]{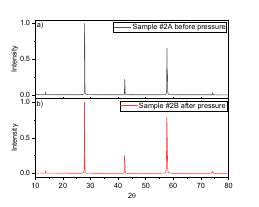}
\caption{\label{fig:figS8} Single-crystal X-ray diffraction results of Sample~\#2 (a) before and (b) after the pressure cycle, measured on the \textit{c}-plane. }
\end{figure*}

Single-crystal X-ray diffraction (XRD) measurements for Sample~\#2 were performed on the \textit{c}-plane, as shown in Fig.~\ref{fig:figS8}. Based on five diffraction peaks, the extracted \textit{c}-axis lattice constants before and after the pressure cycle are $12.793 \pm 0.034$~\AA\ and $12.802 \pm 0.031$~\AA, respectively. These results indicate no measurable change in the lattice constant within the experimental uncertainty.

\begin{figure*}[h!]
\centering
\includegraphics[width=1\linewidth]{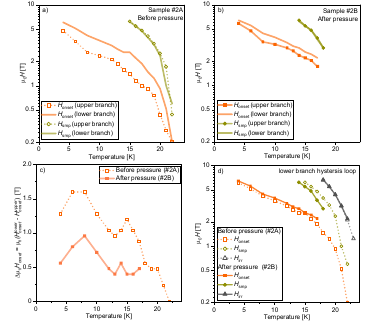}
\caption{\label{fig:figS9} Comparison of $H_{\mathrm{smp}}$ and $H_{\mathrm{onset}}$ extracted from the upper and lower branches of the hysteresis loop for Sample~\#2 for (a) before and (b) after the pressure cycle. (c) Difference in onset field between the upper and lower branches, $\Delta H_{\mathrm{onset}}$, before and after the pressure cycle. (d) Temperature dependence of the lower-branch $H_{\mathrm{onset}}$ and $H_{\mathrm{smp}}$ for Sample~\#2 before and after the pressure cycle.}
\end{figure*}

\begin{figure*}[h!]
\centering
\includegraphics[width=1\linewidth]{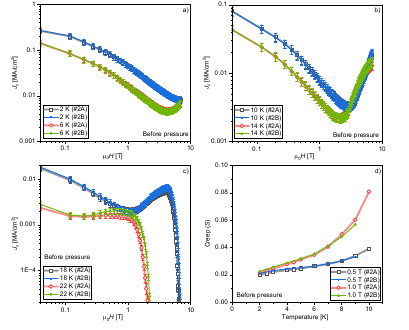}
\caption{\label{fig:figS10} Comparison of $J_c(B)$ for Samples~\#2A and \#2B before the pressure cycle at (a) 2 and 6 K, (b) 10 and 14 K, and (c) 18 and 22 K, showing that the two samples are consistent with each other within the experimental uncertainty. (d) Comparison of $S(T)$ for Samples~\#2A and \#2B before the pressure cycle at 0.5 and 1 T, showing that the differences are within 6\%. }
\end{figure*}

\begin{table}[h!]
\caption{Sample dimensions.}
\label{tab:samples1}
\begin{center}
\begin{tabular*}{0.9\textwidth}{@{\extracolsep{\fill}}ccccccc}
\hline\hline
sample & after Cycle & length $l$ & width $w$ & thickness $\delta$ \\
\# & status & mm & mm & $\mu$m \\
\hline
1 & after 1 pressure cycle & $0.981\pm0.231$ & $0.93\pm0.162$ & $46.12\pm3.65$ \\
2A & before pressure cycle & $0.968\pm0.089$ & $0.798\pm0.058$ & $74.70\pm3.10$ \\
2B & before pressure cycle & $1.283\pm0.077$ & $1.068\pm0.069$ & $74.78\pm8.21$ \\
2B & after 1 pressure cycle & $1.229\pm0.122$ & $0.997\pm0.157$ & $57.17\pm6.78$ \\

\hline\hline
\end{tabular*}
\end{center}
\end{table}

\end{document}